\documentclass{llncs}

\newif\ifpdf
\ifx\pdfoutput\undefined
 \pdffalse
 \usepackage{graphicx}
\else
 \usepackage[pdftex]{graphicx}
 \usepackage{times}
 \pdftrue
 \pdfoutput=1          % output PDF
\fi

\newcommand{\pict}[6]{%
 \begin{figure}[!ht]\begin{center}
  \ifpdf
	\includegraphics[#3]{#4}
  \else
	\includegraphics[#1]{#2}
  \fi
  \caption{#5}
  \label{#6}
 \end{center}\end{figure}
}

 \title{Pyramidal Clustering Algorithms in ISO--3D Project}

 \author{Oldemar Rodr\'{\i}guez\inst{1} \and
	Edwin Diday\inst{1}}

\institute{University Paris 9 Dauphine,\\
	Ceremade. Pl. Du Ml de L. de Tassigny. 75016\\
	\email{orodrigu@ceremade.dauphine.fr}  \\
	\email{diday@ceremade.dauphine.fr}
}

 \titlerunning{Pyramidal Clustering Algorithms}  % abbreviated title (for running head)
 \authorrunning{Oldemar Rodr\'{\i}guez et al.}   % abbreviated author list (for running head)
\index{Rodr\'{\i}guez, O.}
\index{Diday, E.}

\tocauthor{Oldemar Rodr\'{\i}guez (Dauphine, Paris),
  Edwin Diday (Dauphine, Paris)}

\date{\today}

\begin{document}
\ifpdf\DeclareGraphicsExtensions{.pdf,.png,.jpg,.mps}\else
		\DeclareGraphicsRule{.mps}{eps}{.mps}{}
		\DeclareGraphicsRule{.png}{bmp}{}{}
		\DeclareGraphicsRule{.gif}{bmp}{}{}
		\DeclareGraphicsRule{.tif}{bmp}{}{}
\fi

\maketitle
\begin{abstract}
\noindent
Pyramidal clustering method generalizes hierarchies by allowing non-disjoint
classes at a given level instead of a partition. Moreover, the clusters of
the pyramid are intervals of a total order on the set being clustered.
[Diday 1984], [Bertrand, Diday 1990] and [Mfoumoune 1998] proposed
algorithms to build a pyramid starting with an arbitrary order of the
individual. In this paper we present two new algorithms name {\tt CAPS} and
{\tt CAPSO}. {\tt CAPSO} builds a pyramid starting with an order given on
the set of the individuals
(or symbolic objects) while {\tt CAPS} finds this order.
These two algorithms allows moreover to cluster more
complex data than the tabular model allows to process, by considering
variation on the values taken by the variables, in this way, our method
produces a symbolic pyramid. Each cluster thus formed is defined not only by
the set of its elements (i.e. its extent) but also by a symbolic object,
which describes its properties (i.e. its intent). These two algorithms were implemented
in C++ and Java to the ISO--3D project.
\end{abstract}

\section{Definitions}

Diday in \cite[Diday (1984)]{Kn:Did1} proposes the algorithm {\sf CAP}\ to
build numeric pyramids. Algorithms are also presented with this purpose in
\cite[Bertrand y Diday (1990)]{Kn:Ber2}, \cite[Gil (1998)]{Kn:Gil} and
\cite[Mfoumoune (1998)]{Kn:Mfo}. Paula Brito in \cite[Brito (1991)]{Kn:Bri1}
proposes a macro--algorithm that generalizes the algorithm to build numeric
pyramids proposed by Bertrand to the symbolic case. In this article we
propose two algorithm designed to build symbolic pyramids ({\sf CAPS} and {\tt CAPSO}), that
is to say, a pyramid in which each node is again a symbolic object. These
algorithms also calculate the extension of each one of these symbolic
objects and verifie its completeness.

{\bf Notation:}

\begin{itemize}
\item  $\Omega $ the set of individuals.

\item  $O_j$ the description space for the variable $j$.

\item  $P(O_j)$ the set of parts of $O_j$.

\item  The description of an individual $\omega $ is represented by the
vector \linebreak $(y_1(\omega ),\ldots ,y_p(\omega ))$ where each variable $y_j$, $%
j=1,2,\ldots ,p$ is an application of $\Omega $ in $P(O_j)$. The value of $%
y_j(\omega )$ can be represented by a group of values, an interval or a
histogram, among others.

\item  Let $D=P(O_1)\times P(O_2)\times \cdots \times P(O_p)$ the set of the
possible descriptions and $d\in D$ a description, so for every $%
j=1,2,\ldots ,p$, $d_j$ represents a description like a set of values.
\end{itemize}

In \cite[Diday (1999)]{Kn:Did5} the following definition of Symbolic Object
is presented:

\begin{definition}
A symbolic object is a triple $(a,R,d)$ where $R$ in a vector of
relationships $R_i$, $d=(d_1,d_2,\ldots ,d_p)$ is a vector of descriptions $%
d_i$, and $a$ is an application of $\Omega $ in $\{T,F\}$.
\end{definition}

If in the previous definition we take $a(w)=[y_1(w)R_1d_1]\wedge
[y_2(w)R_2d_2]\wedge \cdots \wedge [y_p(w)R_pd_p]$ where $a(w)=T$ iff $%
y_j(w)R_jd_j$ for all $j=1,2,\ldots ,p$ then the symbolic object is known
like Object of Assertion.

If $[y_j(w)R_jd_j]\in L=\{T,F\}$ for all $j=1,2,\ldots ,p$ the symbolic
object is known like {\it Boolean Object} and if $[y_j(w)R_jd_j]\in L=[0,1]$
for all $j=1,2,\ldots ,p$ the symbolic object is known as{\it \ Modal Objet}%
.

In the case of boolean objects the extent is define for $ext_\Omega
(a)=\{w\in \Omega $ such that $a(w)=T\}$; while in the case of modal
symbolic objects the extent of $a$ in the level $\alpha $ is defined for $%
ext_\Omega (a,\alpha )=\{w\in \Omega $ such that $a(w)\geq \alpha \}$.

For the construction of Symbolic Pyramids will be necessary to calculate
the union among symbolic objects, this operation is
defined like it continues \cite[Diday (1987)]{Kn:Did3}.

\begin{definition}
Let $s_1=(a_{1,}R,d_1)$ and $s_2=(a_{2,}R,d_2)$ two symbolic objects, the
union between $s_1$ y $s_2$ denoted for $s_1\cup s_2,$ is defined as the
union of all the symbolic objects $e_i$ such that for all $i$ we have that $%
ext_\Omega (e_i)\supseteq ext_\Omega s_1\cup ext_\Omega s_2$.
\end{definition}

An important concept inside the symbolic pyramidal classification is the
completeness of the symbolic object. A symbolic object is
complete if it describes in an exhaustive way its extension, A formal
definition is presented it is {\rm \cite[Brito (1991)]{Kn:Bri1}}.

\begin{definition}
{\rm \label{GraGen} }Let $s=\bigwedge\limits_{j=1}^pe_j$ a symbolic object,
the Degree of Generality of $s$ is defined by:
\[
G(s)=\prod_{j=1}^p G(e_j)
\]
where
\[
G(e_j)=\left\{
\begin{array}{lll}
\frac{|V_j|}{|{\cal Y}_j|} & \mbox{if} & e_j=[y_j\in V_j]\mbox{, }%
V_j\subseteq {\cal Y}_j\mbox{ with }{\cal Y}_j\mbox{ discreet.} \\
\frac{\mbox{length}(V_j)}{\mbox{length}({\cal Y}_j)} & \mbox{if} &
e_j=[y_j\in V_j]\mbox{, }V_j\subseteq {\cal Y}_j\mbox{ with }{\cal Y}_j\mbox{
continuous.} \\
\frac{\sum_{h=1}^kw_h^{}}k & \mbox{if} &
\begin{array}{l}
e_j=[y_j=\{m_1(w_1),\ldots ,m_k(w_k)\}]\mbox{ is a distribution of} \\
\mbox{frequency of the variable }y_j\mbox{ discreet.}
\end{array}
\end{array}
\right.
\]
\end{definition}

%         L I S T A   D E   C O M A N D O S   P A R A   T e X

\def\empoint#1{\special{em:point #1}}               
\def\empiramide#1#2#3#4{\special{em:line #1, #2}\special{em:line #2, #3}\special{em:line #3, #4}}
\def\LATEX{L\kern-.3em\raise.8ex\hbox{a}\TeX}
\newcommand{\E}{\mbox{$ \backslash $}}
\newcommand{\R}{\mbox{$ I \hspace{-1.2mm} R $}}
\newcommand{\Rn}{\mbox{$ I \hspace{-1.2mm} R^{n} $}}
\newcommand{\N}{\mbox{$ I \hspace{-1.2mm} N $}}
\newcommand{\K}{\mbox{$ I \hspace{-1.2mm} K $}}
\newcommand{\C}{\mbox{\rule[.15mm]{.25mm}{2.88mm}\hspace{2.8mm} C}}
\newcommand{\ZZ}{\mbox{$ Z \hspace{-2.2mm} Z $}}
\newcommand{\SK}{\mbox{$ {\cal S}{\cal I}{\cal D}{\cal E}{\cal K}{\cal I}
{\cal C}{\cal K} $}}
\newcommand{\X}{\mbox{$ {\tt X} $}}
\newcommand{\Y}{\mbox{$ {\tt Y} $}}
\newcommand{\W}{\mbox{$ {\tt W} $}}
\newcommand{\V}{\mbox{$ {\tt V} $}}
\newcommand{\A}{\mbox{$ {\tt A} $}}
\newcommand{\Z}{\mbox{$ {\tt Z} $}}
%\newcommand{\f}{\mbox{$ \tilde{f} $}}
%newcommand{\g}{\mbox{$ \tilde{g} $}}
\newcommand{\ii}{\'{\i}}
\newcommand{\FT}{\mbox{$ \tilde{{\cal F}} $}}
\newcommand{\PT}{\mbox{$ \hspace{5mm} \forall \hspace{3mm} $}}
\newcommand{\exis}{\mbox{$ \hspace{3mm} \exists \hspace{2mm} $}}
\newcommand{\sii}{\mbox{$ \hspace{3mm} \Longleftrightarrow \hspace{2mm} $}}
\newcommand{\I}{\mbox{$ I \hspace{-1.4mm} I $}}
\newcommand{\flecha}{\mbox{$ \longrightarrow $}}
\newcommand{\ve}[1]{\overrightarrow{#1}}
\newcommand{\gfrac}[2]{\displaystyle{\frac{#1}{#2}}}
\newcommand{\gsum}[2]{\mbox{$ \displaystyle{\sum_{#1}^{#2}} $}}
\newcommand{\gint}{\mbox{$\displaystyle{\int}$}}
\newcommand{\gprod}[2]{\mbox{$ \displaystyle{\prod_{#1}^{#2}} $}}

\newcommand{\intd}[2]{\gint \!\! \gint_{\!\!\! #1} {#2} \: dxdy}

  % T E S I S

\newcommand{\tq}{\;\;\; | \;\;\;}
\newcommand{\mn}[1]{ min \raisebox{-3mm}{\footnotesize{\hspace{-9mm}$#1$\hspace{3mm}}}}
\newcommand{\mnl}[1]{ min \raisebox{-3mm}{\footnotesize{\hspace{-5mm}$#1$\hspace{3mm}}}}
\newcommand{\ci}{\Omega \raisebox{3.5mm}{\hspace{-2mm}$\circ$}}
\newcommand{\inte}[1]{#1 \raisebox{3.5mm}{\hspace{-3.5mm}$\circ$}}
\newcommand{\xp}{x^{\prime}}
\newcommand{\xt}{\widehat{x}}
\newcommand{\gr}[1]{\nabla f(#1)}
\newcommand{\g}{\nabla f(x)}
\newcommand{\gl}{\nabla L(x)}
\newcommand{\p}{P_{I^{\prime}}}
\newcommand{\f}{F_{I^{\prime}}}
\newcommand{\po}[1]{P_{#1}^{\bot}}
\newcommand{\lo}[1]{L_{#1}^{\bot}}
\newcommand{\Md}[4]{\left(\begin{array}{rr}
                           #1  &  #2  \\
                           #3  &  #4
                          \end{array}
                     \right) }
\newcommand{\Vd}[2]{\left(\begin{array}{r}
                           #1  \\  #2
                          \end{array}
                     \right) }
\newcommand{\Mt}[9]{\left(\begin{array}{rrr}
                           #1  &  #2  & #3 \\
                           #4  &  #5  & #6 \\
                           #7  &  #8  & #9
                          \end{array}
                     \right) }
\newcommand{\Vt}[3]{\left(\begin{array}{r}
                           #1  \\  #2  \\ #3
                          \end{array}
                     \right) }
\newcommand{\Mtd}[6]{\left(\begin{array}{rrr}
                           #1  &  #2  \\
                           #3  &  #4 \\
                           #5  &  #6
                          \end{array}
                     \right) }
\newcommand{\Mdt}[6]{\left(\begin{array}{rrr}
                           #1  &  #2 & #3  \\
                           #4  &  #5 & #6
                          \end{array}
                     \right) }
\newcommand{\Ft}{F_{I_{0}(x)}^{t}}
\newcommand{\F}{F_{I_{0}(x)}}
\newcommand{\Pt}{P_{I_{0}(x)}^{\bot}}
\newcommand{\Ftp}[1]{F_{I_{0}(x)-#1}^{t}}
\newcommand{\Fp}[1]{F_{I_{0}(x)-#1}}
\newcommand{\Ptp}[1]{P_{I_{0}(x)-#1}^{\bot}}

% ------------------------------------------------------

\newcommand{\fdp}[4]{f(x) = \left\{ \begin{array}{ll}
                                      #1  &   \mbox{ #2 }  \\
                                      #3  &   \mbox{ #4 }
                                    \end{array}
                            \right.  }
\newcommand{\fdg}[4]{g(y) = \left\{ \begin{array}{ll}
                                      #1  &   \mbox{ #2 }  \\
                                      #3  &   \mbox{ #4 }
                                    \end{array}
                            \right.  }

\newcommand{\fdh}[4]{h(y) = \left\{ \begin{array}{ll}
                                      #1  &   \mbox{ #2 }  \\
                                      #3  &   \mbox{ #4 }
                                    \end{array}
                            \right.  }

\newcommand{\fda}[4]{F(x) = \left\{ \begin{array}{ll}
                                      #1  &   \mbox{ #2 }  \\
                                      #3  &   \mbox{ #4 }
                                    \end{array}
                            \right.  }

\newcommand{\fdpt}[6]{f(x) = \left\{ \begin{array}{ll}
                                      #1  &   \mbox{ #2 }  \\
                                      #3  &   \mbox{ #4 }  \\
                                      #5  &   \mbox{ #6 }
                                    \end{array}
                            \right.  }

\newcommand{\fdpg}[6]{g(x) = \left\{ \begin{array}{ll}
                                      #1  &   \mbox{ #2 }  \\
                                      #3  &   \mbox{ #4 }  \\
                                      #5  &   \mbox{ #6 }
                                    \end{array}
                            \right.  }

\newcommand{\fd}[4]{f(x,y) = \left\{ \begin{array}{ll}
                                      #1  &   \mbox{ #2 }  \\
                                      #3  &   \mbox{ #4 }
                                    \end{array}
                            \right.  }

Diday generalizes in \cite[Diday (1984)]{Kn:Did1} the concept of binary
hierarchy to the pyramid, like we show in the following definitions.

\begin{definition}
\begin{itemize}
\item  Let be $\theta $ a total order on $\Omega $ and $P$ a set of parts
not empty of $\Omega $. An element $h\in P$ is said connected according to
the total order $\theta $, if for every $w\in \Omega $ that is between the $%
\max (h)$ and the $\min (h)$ ($\min (h)\theta w\theta \max (h)$) we have
that $w\in h$.

\item  A total order $\theta $ on $\Omega $ is compatible with $P,$ the set
of parts of $\Omega $, if all element $h\in P$ is connected according to the
total order $\theta $.
\end{itemize}
\end{definition}

\begin{definition}
Let be $\Omega $ a finite set and $P$ a set of parts not empty of $\Omega $
(called nodes), $P$ is a pyramid if it has the following properties:%
\vspace{-1mm}

\begin{enumerate}
\item  $\Omega \in P$.

\item  $\forall $ $w\in \Omega $ we have that $\{w\}\in P$ (terminal nodes).

\item  $\forall $ $(h,h^{^{\prime }})\in P\times P$ we have that $h\cap
h^{^{\prime }}\in P$ or $h\cap h^{^{\prime }}=\emptyset $.

\item  A total order exists $\theta $ in $\Omega $ compatible with $P$.
\end{enumerate}
\end{definition}

\begin{definition}
$\label{PirSim}$Let be $\Omega $\ a finite set of symbolic objects and let
be $P$\ a set of parts not empty of $\Omega $\ (calls also nodes), $P$\ is a
symbolic pyramid if it satisfies the following properties:\vspace{-1mm}

\begin{enumerate}
\item  $P$\ is a pyramid.

\item  Each node of $P$\ has associate a complete symbolic object.
\end{enumerate}
\end{definition}

Subsequently we present the necessary definitions for the specification of
the algorithm, these definitions differ a little to the definitions
presented in (\cite[Brito (1991)]{Kn:Bri1}, \cite[Bertrand and Diday (1990)]
{Kn:Ber2} and \cite[Mfoumoune (1998)]{Kn:Mfo}), because all are local to the
``component connected''.

For the following definitions we consider a set ${\cal P}$ $\subseteq
P(\Omega )$ (the set of parts of $\Omega $) that it is not necessarily a
pyramid, is possibly a ``pyramid under construction'', for abuse of the
language we will denominate like a node all the element of ${\cal P}$.

\begin{definition}
\begin{itemize}
\item  Let be $C\in {\cal P}$, $C$ is called {\it connected component} if a
total order exists $\leq _C$ associated to $C$.

\item  A node $G\in {\cal P}$ {\it belongs to a connected component} $C$ of $%
{\cal P}$ if $G\subseteq C$. We will also say that the total order $\leq _C$
associated to $C$ induces a total order $\leq _G$ on $G$ in the following
way, if $x,y\in G$ then $x\leq _Gy\Leftrightarrow x\leq _Cy$.

\item  Let be $G_1$ and $G_2$ nodes of ${\cal P}$. Then $G_1$ {\it is
interior} $G_2$ if:\vspace{-1mm}

\begin{itemize}
\item  $G_1\neq G_2$.

\item  $G_1$ and $G_2$ belong to the same connected component $C$.

\item  $\min (G_2)<_C\min (G_1)$ y $\max (G_1)<_C\max (G_2)$, where $\alpha
<_C\beta $ means that $\alpha \leq _C\beta $ and $\alpha \neq \beta $.
\end{itemize}

\item  Let be $G_1$ and $G_2$ nodes of ${\cal P}$, then $G_1$ is {\it %
successor} $G_2$ ($G_2$ is {\it predecessor} $G_1$) if:

\begin{itemize}
\item  $G_1\subset G_2$ in strict sense.\vspace{-1mm}

\item  Doesn't exist a node $G\in {\cal P}$ such that $G_1\subset G\subset
G_2$ in strict sense.
\end{itemize}

\item  A node $G\in {\cal P}$ is called {\it maximal} if it doesn't have
predecessors.

\item  Let be $G_1$ and $G_2$ nodes of ${\cal P}$, then $G_1$ is to the {\it %
left} of $G_2$ ($G_2$ is to the {\it right} of $G_1$) if:\vspace{-1mm}

\begin{itemize}
\item  Belong to the same connected component $C$.

\item  $\min (G_1)\leq _C\min (G_2)$ and $\max (G_1)\leq _C\max (G_2)$.
\end{itemize}

\item  Let be $G_1$ and $G_2$ nodes of ${\cal P}$, then $G_1$ is {\it %
strictly to the left} of $G_2$ if:\vspace{-1mm}

\begin{itemize}
\item  Belong to the same connected component $C$.

\item  $\min (G_1)<_C\min (G_2)$ and $\max (G_1)=\max (G_2)$.
\end{itemize}

\item  Let be $G_1$ and $G_2$ nodes of ${\cal P}$, the $G_2$ is {\it %
strictly to the right} of $G_1$ if:\vspace{-1mm}

\begin{itemize}
\item  Belong to the same connected component $C$.

\item  $\min (G_1)=\min (G_2)$ and $\max (G_1)<_C\max (G_2)$.
\end{itemize}

\item  Let be $G_1$ and $G_2$ nodes of ${\cal P}$, then $G_1$ it is the {\it %
maximal left node} of $G_2$ if:\vspace{-1mm}

\begin{itemize}
\item  $G_1$ is to the left of $G_2$.

\item  $G_1$ is a maximal node.

\item  $\max (G_2)=\max (G_1)$.
\end{itemize}

\item  Let be $G$ a node of ${\cal P}$ that it belongs to the connected
component $C$, let $G_{1,}G_2,\ldots ,G_l$, all the maximal nodes of the
connected component $C$, order of left to right with the order $\leq _C$
(that is to say $G_i$ is to the left of $G_{i+1}$). If $G_m$ is the left
maximal node of $G$ and $m<l$ then $G_{m+1}$ is called the {\it next node
maximal of} $G$. If $m=l$ then $G$ doesn't have {\it next node maximal}.
\end{itemize}
\end{definition}

\begin{definition}[Aggregation conditions]
Let be $G_1$ and $G_2$ two nodes of ${\cal P}$. \vspace{-1mm}

\begin{description}
\item[Case 1:]  If $G_1$ and $G_2$ belong to the same connected component
and we denote for $\overleftarrow{G}$ the left maximal node of $G_1$ and by $%
\overrightarrow{G}$ the right maximal node of $G_1$ (if it exists). Then $G_1
$ and $G_2$ they are aggregables if the two following conditions are
satisfied: \vspace{-1mm}

\begin{enumerate}
\item  $G_1$ is to the right of $\overleftarrow{G}$ and strictly to the left
of $\overleftarrow{G}\cap \overrightarrow{G}$.

\item  $G_2$ is to the left of $\overrightarrow{G}$ and strictly to the
right of $\overleftarrow{G}\cap \overrightarrow{G}$.
\end{enumerate}

\item[Case 2:]  If $G_1$ and $G_2$ do NOT belong to the same connected
component and we denote for $C_1$ and $C_2$ the connected components that $%
G_1$ and $G_2$ belong respectively. Then $G_1$ and $G_2$ are aggregables if
the two following conditions are satisfied: \vspace{-1mm}

\begin{enumerate}
\item  $\min (G_1)=\min (C_1)$ or $\max (G_1)=\max (C_1)$.

\item  $\min (G_2)=\min (C_2)$ or $\max (G_2)=\max (C_2)$.
\end{enumerate}
\end{description}
\end{definition}

\begin{definition}
A node $G$\ of ${\cal P}$\ is called {\it active} if the following three
conditions are satisfied:\vspace{-1mm}

\begin{itemize}
\item  Another node exists $G^{\star }$\ in ${\cal P}$\ such that $G$\ is
aggregated with $G^{\star }$.

\item  $\not \exists \widetilde{G}\in {\cal P}$\ such that $G$\ it is interior
%\item  $\widetilde{G} \in {\cal P}$\ such that $G$\ it is interior
node of $\widetilde{G}$.

\item  $G$\ has been aggregated at most once (one time).
\end{itemize}
\end{definition}

%\newpage

\section{The algorithms}

{\large {\sc Algorithm -- CAPS}}

\vspace{-2mm}

\begin{description}
\item[Input]  : \linebreak

\vspace{-4mm}

\begin{itemize}
\item  $M=$Maximum number of iterations.

\item  $N=$Number of vectors of symbolic data (number of the rows of the
symbolic data table).

\item  $P=$ Number of variables (number of columns of the symbolic data
table).

\item  $X=$Symbolic data table.
\end{itemize}

\item[Output]  : \linebreak

\vspace{-4mm}

\begin{itemize}
\item  A total order ``$\leq $'' on the set $\Omega $ of objects.

\item  A pyramidal structure, that is to say, a succession of quadruples $%
(p,p_I,p_D,f(p))$, with $p=1,2,\ldots ,NG$, where $NG=$total number of nodes
of the pyramid, $p_I=$left son of the node $p$ and $p_D=$right son of the
node $p$. If $p$ is a terminal node then $p_I=p_D=0$.

\item  A symbolic object $O_p$ associated to the node $p$, with $%
p=1,2,\ldots ,NG$.

\item  The extension of the object associated to each node, that is to say, $%
ext(O_p)$, with $p=1,2,\ldots ,NG$.

\item  If the algorithm fails the exit will be an error message.
\end{itemize}

\item[Step 1:]  Initialization phase

\begin{description}
\item[Step 1.1]  $h=1$, where $h$ is the number of iterations.

\item[Step 1.2]  $NG=N$, where $NG=$total number of nodes of the pyramid.

\item[Step 1.3]  $NC=N$, where $NC=$Number of connected components, in one  
given iteration (at the end of the execution of the algorithm we will have $NC=1$).

\item[Step 1.4]  $NP=N$, where $NP=$number of active nodes in a given
iteration (at the end of the execution of the algorithm we will have $NP=1$).

\item[Step 1.5]  We initialize the $N$ initial quadruples of the pyramidal
structure: $(s,0,0,0)$, $s=1,2,\ldots ,N$.

\item[Step 1.6]  We build: $NC$ initials connected components  $C_s=\{s\}$, $%
s=1,2,\ldots ,NC$, a total order $\leq _C$ associated to each connected
component, at the begining one we have that $s\leq _Cs$. It is also denoted
for $C$ to the set formed by all the components, that is to say, $%
C=\{C_1,C_2,\ldots ,C_{NC}\}$.

\item[Step 1.7]  We build: $NP$ initials nodes active in the following way $%
G_q=\{(\alpha ,\beta ,s_q,\ell )\}$, for $q=1,2,\ldots ,NP$. Where $\alpha $
is the number that is associated to each active node in a given iteration
(the active nodes will be numbered from $1$ to $NP$), $\beta $ is the gloabl number
of the node (the first node generated by the algorithm is $\beta =N+1$%
, the second node generated by the algorithm is $\beta =N+2$ and so on), $s_q
$ is the vector of symbolic data stored in the row $q-$th of the symbolic
data table (the beginning each row of the matrix corresponds a node,
however, when the algorithm advances a node can correspond to the union
of several symbolic objects, that is to say, can be associated to the
``union of several rows of the symbolic data table'') and $\ell $ is the
number of times that the node has been aggregated ($\ell \leq 2$). We denote for $%
G=\{G_s\}_{s=1,2,\ldots ,NP}=\{(1,1,s_1,0),(2,2,s_2,0),\ldots
,(NP,NP,s_{NP},0)\}$ the set of all the initials actives nodes, $%
G_q^1=\alpha $, $G_q^2=\beta $, $G_q^3=s_q$ and $G_q^4=\ell $

\item[Step 1.8]  We calculate the initial matrix of  ``disimilarities'' $%
D_{ij}^h=G(s_i\cup s_j)$ where $s_k$ is the vector of symbolic data stored
in the row $k-$th row of the symbolic data table, with $i,j=1,2,\ldots ,N$.
\end{description}

\item[Step 2:]  Elimination phase

\begin{description}
\item[Step 2.1]  We find the nodes that are aggregables and with which nodes they
are aggregables, that is to say, we calculate the matrix:
\begin{eqnarray*}
B_{lu}=\left\{
\begin{array}{lll}
1 & \mbox{if} & G_l\mbox{ and }G_u\mbox{ are aggregables} \\
0 & \mbox{if} & G_l\mbox{ and }G_u\mbox{ are not aggregables} \\
0 & \mbox{if} & \exists \mbox{ }\widetilde{G}\in {\cal P}\mbox{ such that }%
G_l\mbox{ is an interior node of }\widetilde{G} \\
0 & \mbox{if} & \exists \mbox{ }\widetilde{G}\in {\cal P}\mbox{ such that }%
G_u\mbox{ is an interior node of }\widetilde{G}
\end{array}
\right. \mbox{ } \\
\mbox{for }l,u=1,2,\ldots ,NP\mbox{.}
\end{eqnarray*}

\item[Step 2.2]  We calculate the active nodes that are noy any more
aggregables with any other node (that is to say the nodes that will not be
active), that is to say, we find all the nodes $G\eta $ such that, the row
and the column $\eta $ of the matrix $B$ contain only zeros. Let $\widetilde{%
G}=\{G_{\alpha _1},G_{\alpha _2},\ldots ,G_{\alpha _m}\}$ with $m\geq 0$ the
set of those $m$ nodes non aggregables.

\item[Step 2.3]  $NP=NP-m$.

\item[Step 2.4]  $G=G - \widetilde{G}$.

\item[Step 2.5]  Upgrades the matrix of distances $D^h$ so that: \newline
$D^h\in M_{(NP-m)\times (NP-m)}$, because we eliminated of $D^h$ all the
rows and columns associated to nodes not activate.
\end{description}

\item[Step 3:]  Phase of formation of new nodes (Step of Generalization)

\begin{description}
\item[Step 3.1]  We find $s_i$ and $s_j$ such that $D_{ij}^h=G(s_i\cup s_j)$
is minimum and $B_{ij}=1$, where $i,j=1,2,\ldots ,NP$. The nodes where this
minimum is reached is denoted for $s_{i^{\star }}$ and $s_{j^{\star }}$. If $%
B_{ij}=0,$ $\forall $ $i,j=1,2,\ldots ,NP$ then the algorithm finishes and
an returns error message, if not continuous in the step 3.2.

\item[Step 3.2]  $NG=N+h$, subsequently we calculate the following
quadruple of the pyramidal structure $(NG,G_{i^{\star }}^2,G_{j^{\star
}}^2,D_{i^{\star }j^{\star }}^h)$.

\item[Step 3.3]  We calculate $s^{\star }=s_{i^{\star }}\cup s_{j^{\star }}$
and its extension $ext(s^{\star })$.

\item[Step 3.4]  If $s^{\star }$ is complete and $ext(s^{\star
})=ext(s_{i^{\star }})\cup ext(s_{j^{\star }})$ then the algorithm
continuous in the step 4, if not we take $B_{i^{\star }j^{\star }}=0$ and
the algorithm returns to the step 3.1.
\end{description}

\item[Step 4:]  Phase of upgrade

\begin{description}
\item[Step 4.1]  $h=h+1$.

\item[Step 4.2]  (Upgrade of the components) If $G_{i^{\star }}^{}\in
C_{\sigma _1}$ and $G_{j^{\star }}^{}\in C_{\sigma _2}$ such that $\sigma
_1\neq \sigma _2$ (belong to different connected components)

\begin{description}
\item[Step 4.2.1]  We forme a new connected component $C_\sigma =C_{\sigma
_1}\cup C_{\sigma _2}$, then in $C_\sigma $ we define a new total order, for
this four possibilities exist:

\begin{description}
\item[Caso 1:]  If $\max (G_{i^{\star }}^{})=\max (C_{\sigma _1})$ and $\min
(G_{j^{\star }}^{})=\min (C_{\sigma _2})$ :\newline
If $x,y\in C_\sigma $ then $x\leq _{C_\sigma }y\Leftrightarrow \left\{
\begin{array}{lll}
x\leq _{C_{\sigma _1}}y & \mbox{if} & x,y\in C_{\sigma _1} \\
x\leq _{C_{\sigma _2}}y & \mbox{if} & x,y\in C_{\sigma _2} \\
x\in C_{\sigma _1}\mbox{ } & \mbox{and} & y\in C_{\sigma _2}
\end{array}
\right. $

\item[Caso 2:]  If $\max (G_{i^{\star }}^{})=\max (C_{\sigma _1})$ and $\max
(G_{j^{\star }}^{})=\max (C_{\sigma _2})$: \newline
If $x,y\in C_\sigma $ then $x\leq _{C_\sigma }y\Leftrightarrow \left\{ 
\begin{array}{lll}
x\leq _{C_{\sigma _1}}y & \mbox{if} & x,y\in C_{\sigma _1} \\ 
y\leq _{C_{\sigma _2}}x & \mbox{if} & x,y\in C_{\sigma _2} \\ 
x\in C_{\sigma _1}\mbox{ } & \mbox{and} & y\in C_{\sigma _2}
\end{array}
\right. $

\item[Caso 3:]  If $\min (G_{i^{\star }}^{})=\min (C_{\sigma _1})$ and $\min
(G_{j^{\star }}^{})=\min (C_{\sigma _2})$: \newline
If $x,y\in C_\sigma $ then $x\leq _{C_\sigma }y\Leftrightarrow \left\{ 
\begin{array}{lll}
y\leq _{C_{\sigma _1}}x & \mbox{if} & x,y\in C_{\sigma _1} \\ 
x\leq _{C_{\sigma _2}}y & \mbox{if} & x,y\in C_{\sigma _2} \\ 
x\in C_{\sigma _1}\mbox{ } & \mbox{and} & y\in C_{\sigma _2}
\end{array}
\right. $

\item[Caso 4:]  If $\min (G_{i^{\star }}^{})=\min (C_{\sigma _1})$ and $\max
(G_{j^{\star }}^{})=\max (C_{\sigma _2})$: \newline
If $x,y\in C_\sigma $ then $x\leq _{C_\sigma }y\Leftrightarrow \left\{ 
\begin{array}{lll}
y\leq _{C_{\sigma _1}}x & \mbox{if} & x,y\in C_{\sigma _1} \\ 
y\leq _{C_{\sigma _2}}x & \mbox{if} & x,y\in C_{\sigma _2} \\ 
x\in C_{\sigma _1}\mbox{ } & \mbox{and} & y\in C_{\sigma _2}
\end{array}
\right. $
\end{description}

\item[Step 4.2.2]  $NC=NC-1$.

\item[Step 4.2.3]  $C=(C - \{C_{\sigma _1},C_{\sigma _2}\})\cup
\{C_\sigma \}$.
\end{description}

\item[Step 4.3]  (Upgrade of the active nodes)

\begin{description}
\item[Step 4.3.1]  The new node is calculated: $G_\sigma =G_{i^{\star }}\cup
G_{j^{\star }}:=\{(G_{i^{\star }}^1,N+h,s_{i^{\star }}\cup s_{j^{\star
}},0)\}$ and we upgrade the number of times that these two nodes have been
aggregated, that is to say, $G_{i^{\star }}^4=G_{i^{\star }}^4+1$ and $%
G_{j^{\star }}^4=G_{j^{\star }}^4+1$. Then the nodes that have been
aggregated twice are eliminated, for this four possibilities exist:

\begin{description}
\item[Caso 1:]  If $G_{i^{\star }}^4=2$ and $G_{j^{\star }}^4=2$ (both nodes
have been aggregated twice) then: $NP=NP-1$ and $G=(G -
\{G_{i^{\star }},G_{j^{\star }}\})\cup \{G_\sigma \}$.

\item[Caso 2:]  If $G_{i^{\star }}^4=1$ and $G_{j^{\star }}^4=1$ (both nodes
have been aggregated once) then: $NP=NP+1$ and $G=G\cup \{G_\sigma \}$.

\item[Caso 3:]  If $G_{i^{\star }}^4=2$ and $G_{j^{\star }}^4=1$ ($%
G_{i^{\star }}$ has been aggregated twice and $G_{j^{\star }}$ has been
aggregated once) then: $G=(G - \{G_{i^{\star }}\})\cup
\{G_\sigma \}$.

\item[Caso 4:]  If $G_{i^{\star }}^4=1$ and $G_{j^{\star }}^4=2$ ($%
G_{j^{\star }}$ has been aggregated twice and $G_{i^{\star }}$ has been
aggregated once) then: $G=(G - \{G_{j^{\star }}\})\cup
\{G_\sigma \}$.
\end{description}
\end{description}

\item[Step 4.4]  we calculate the new matrix of ``distances'' $%
D_{ij}^h=G(s_i\cup s_j)$ for $i,j=1,2,\ldots ,NP$.
\end{description}

\item[Step 5:]  If $NP=1$ then the algorithm finishes, otherwise if $h>M$
then the algorithm returns an error message, if not the algorithm return to
the step 2.
\end{description}

The following algorithm allows to build a symbolic pyramid when the order of
the objects is known a priori. This algorithm is in fact a case particulier
of the previous algorithm, since the previous algorithm begins with $%
n=|\Omega |$ connected components, while the following algorithm begins with 
$n=1$.

\vspace{5mm}

{\large {\sc Algorithm (CAPSO)}}

\begin{description}
\item[Input]  : \linebreak

\vspace{-4mm}

\begin{itemize}
\item  The same input of the previous algorithm plus

\item  A total order ``$\leq _\Omega $'' on the set $\Omega $ of objects.
\end{itemize}

\item[Output]  : \linebreak

\vspace{-4mm}

\begin{itemize}
\item  The same input of the previous algorithm except the total order
\end{itemize}

\item[Step 1:]  Initialization phase (It only changes the step 1.6)

\begin{description}
\item[Step 1.6]  A connected component is built $C=\{s_1,s_2,\ldots ,s_N\}$,
with total order $\leq _C$, defined as it continues: $s_i\leq
_Cs_j\Leftrightarrow s_i\leq _\Omega s_j$.
\end{description}

\item[Step 2:]  Elimination phase (Identical to the previous algorithm)

\item[Step 3:]  Phase of formation of new nodes (Step of Generalization)
(Identical to the previous algorithm)

\item[Step 4:]  Phase of upgrade (Identical to the previous algorithm)
\end{description}

\begin{theorem}
The algorithm CAPS builds a symbolic pyramid.
\end{theorem}

\begin{corollary}
The algorithm CAPSO builds a symbolic pyramid.
\end{corollary}

{\bf Example:}

{\scriptsize 
\[
X=\left[ 
\begin{array}{ccccc}
\lbrack 1,4] & 2 & (1(0.4),2(0.1),3(0.2),4(0.07),5(0.02)) & (1(0.1),2(0.9))
& (1(0.7),2(0.2)) \\ 
\lbrack 1,4] & 3 & (1(0.6),2(0.1),3(0.1),5(0.0)) & (1(0.1),2(0.9)) & 
(1(0.7),2(0.2)) \\ 
\lbrack 1,5] & 2 & (1(0.7),2(0.2)) & (1(0.0),2(0.9)) & (1(0.7),2(0.2)) \\ 
\lbrack 1,4] & 1 & (1(0.7),2(0.0),3(0.1),4(0.0),5(0.0),6(0.0),7(0.0)) & 
(1(0.0),2(0.9)) & (1(0.7),2(0.2)) \\ 
\lbrack 1,4] & 1 & (1(0.4),3(0.4),4(0.0),5(0.0)) & (1(0.0),2(0.9)) & 
(1(0.8),2(0.1)) \\ 
\lbrack 1,6] & 2 & (2(0.4),3(0.1),4(0.3),5(0.0),6(0.0),7(0.0)) & 
(1(0.0),2(0.9)) & (1(0.7),2(0.2))
\end{array}
\right] 
\]
}

\pict{width=100mm}{pyr1.eps}{width=100mm}{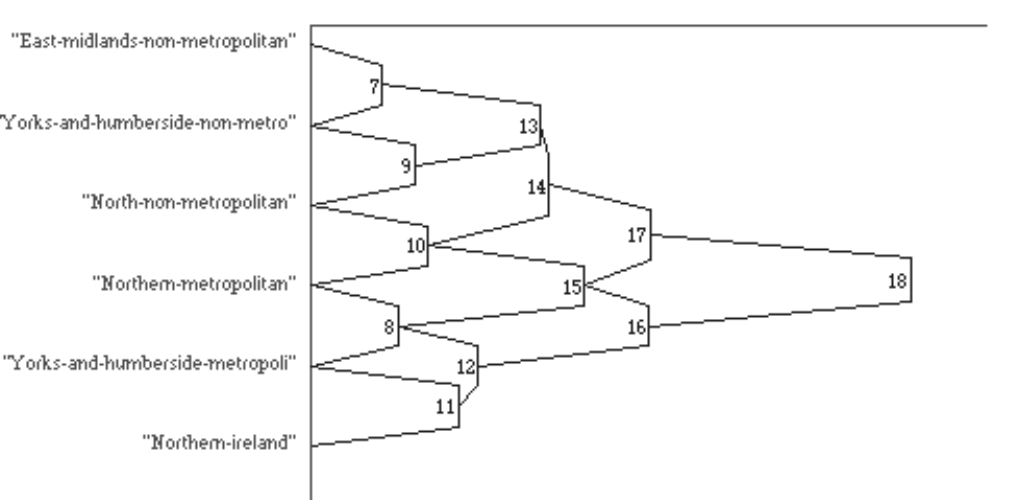}{Pyramid produced using {\tt CAPS}}{fig1}

The algorith {\tt CAPS}, with the matrix $X$ as input, produces the pyramid
of the Figure 1, and if we use the algorithm {\tt CAPSO} with the order 8,
6, 4, 2, 5, 3, 1 given a priori, then we get the pyramid of the Figure 2.

Where the symbolic objects asociated to each node and their extension of the
pyramid presented Figure 1 are:

\begin{verbatim}
P7=[y1=[1.000,4.000]]^[y2={1.00}]^[y3=(1(0.7181),2(0.0537),3(0.4348),
	4(0.0870),5(0.0435),6(0.0134),7(0.0067))]^[y4=(1(0.0435),2(0.9799
	))]^[y5=(1(0.8696),2(0.2483))]
Ext(P7)={4,5}
P8=[y1=[1.000,5.000]]^[y2={2.00}]^[y3=(1(0.7882),2(0.1151),3(0.2806),
	4(0.0791),5(0.0288),6(0.0000),7(0.0000))]^[y4=(1(0.0588),2(0.9856)
	)]^[y5=(1(0.7765),2(0.2734))]
Ext(P8)={1,3}
etc....
\end{verbatim}

\pict{width=100mm}{pyr2.eps}{width=100mm}{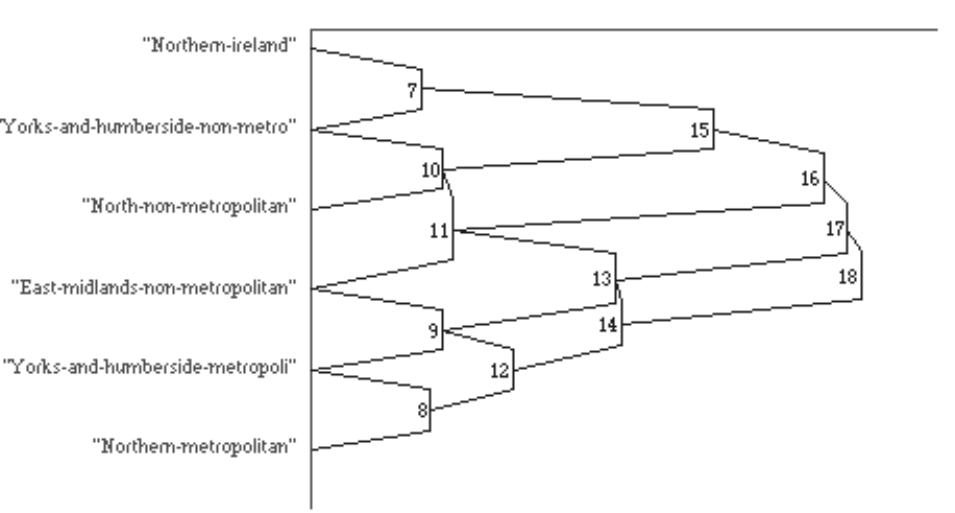}{Pyramid produced by {\tt CAPSO}}{fig2}

\end{document}